# TOWARDS EFFICIENT SERVICE DELIVERY: THE ROLE OF WORKFLOW SYSTEMS IN PUBLIC SECTOR IN KENYA.

**BY: Lawrence Xavier Waweru Thuku and Karanja Evanson Mwangi**

thukulawrence@yahoo.com

**Abstract**

Current challenges in Electronic Government initiatives include performance management, effective and efficient means of sharing information between different stakeholders e.g. government departments in order to improve the quality of service delivery to the citizens. To address these challenges, this paper addresses an application domain of workflow systems in public institutions and proposes a framework that can be effectively used in implementing workflow systems in public institutions.

## Introduction

Majority of the government organizations, are faced with two main competing pressures: The need to deliver high-quality services to citizens and businesses, and at the same time, improve efficiencies and reduce costs. As opposed to the government organizations way of meeting the citizens and businesses needs, the private sector has revolutionized and automated the provision of customer service. For example, in the banking sector, customers are able to access customer support services anytime anywhere, through a range of channels that include access to online information, around-the-clock contact centers, fully transactional Web sites, and interactive Web chat sessions, Short message Notification. This means customers now expect companies to know who they are when they call and to respond to inquiries quickly, track orders precisely, and resolve complaints efficiently. These high standards of customer service in the private sector have in turn raised the expectations of citizens, businesses, and contractors when they deal with government organizations. Citizens and businesses now expect more seamless, personalized, and convenient self-service options for interacting with government organizations, across multiple channels, regardless of time and location.

In the light of these expectations, government institutions are adopting a *citizen-centric* approach for service delivery, inquiry management, and other business practices. They are redesigning their internal business processes to focus on the needs of their citizens, businesses, and other service users. This citizen-centric approach requires centralized access to detailed information about citizens and services, and many government organizations are using CRM (Customer Relationship Management) solutions to deliver these capabilities. Equipped with integrated portals and contact centers, employees can quickly locate and share information, which lets them build and sustain citizen relationships with the same tools private sector organizations use to manage their customer relationships. Managers and front-line workers can resolve inquiries, learn from feedback, and fix the root causes that create service problems.

This paper uses secondary information to address the application domain of workflow systems in public institutions-local authorities in order to solve the current challenges of performance management, sharing of information between stakeholders and improve the quality of service delivery to the citizens. We propose a framework that can effectively be used in implementing workflow systems in the Local authorities.

**Counties, Constitution and Vision 2030**

The Government of Kenya is comprised of three arms: the Executive, Legislature, and Judiciary. In the advent of a new constitution in Kenya (Kenya, 2010) which was promulgated on 27th August 2010, the Local government which will be known as Counties forms a cornerstone in realization of an effective devolved government and plays a great role in economic and political transformation towards the realization of vision 2030 and beyond (Karanja, 2009) . The economic blueprint known as "Vision 2030" is the roadmap detailing Kenya's move to a middle income economic nation with a sustainable growth rate of 10 percent by year 2030 with anticipated milestones such as creating employment, provision of basic needs to all, wiping out famine, preventable deaths and building a democratic system that respect the rule of law, rights and freedom of every individual and society (NESC) .

The counties are charged with the authority to facilitate industrialization through provision of appropriate infrastructure, operation and maintenance of vital services, taxation and licensing, land planning and development to spur growth in their economies, thus providing employment opportunities.

The Constitution of Kenya (Kenya, 2010), Chapter Eleven highlights the objects, principles and strands of the devolved government specifically the counties. Various articles in Chapter Eleven of constitution of Kenya emphasizes the role of counties to realising good governance , efficient service delivery to the citizens , economic and social prosperity , Article 174 of the Constitution of Kenya states as follows:

*"The objects of the devolution of government are…. (f) To promote social and economic development and the provision of proximate, easily accessible services through Kenya."*

In 2009 , only 40 out of 175 local authorities were self sustaining thus able to provide required services without the intervention of the central government (KAM, 2009). Self Sustainance and efficient service delivery in the county government are recognised as key pillars to the overall success of a devolved government structure, Article 175(b) further states as follows:

*"... County Government shall have reliable sources of revenue to enable them to govern and deliver services effectively ….."*

Article 176(2) of the Constitution of Kenya opines that "Every County Government shall decentralize its functions and the provision of its services to the extent that it is efficient and practicable to do so". The counties have to be innovative to realize these provisions.

**Challenges facing local authorities in Kenya**

Like any other institution in the world, the public administration sector is being criticised constantly by its taxpayers who want to see the efficiency and improvement of delivery of its services and culture just like in the private administration such as: in the banks, insurance companies, and even in the trade centers. People keep wondering why bureaucratic public institutions cannot solve their problems, why they have to visit several institutions in order to get the information, or why they have to make so many phone calls to have their problems solved.

Mitullah & Waema (Mitullah & Waema, 2005) supports this citing (Kenya, 1999) on the number of challenges that are facing LAs (Local Authorities) in realizing their mandate here in Kenya. These have in turn resulted into poor service provision and management and many analysts have criticized the LAs, and questioned their role in local development. The challenges included; delivery of infrastructure and services, financial management, institutional and legal framework, human resource capacity and managing rapid growth.

Therefore, for the Kenyan government to be able to achieve the vision 2030 and enhance the efficiency and improve on delivery of its services and culture, the driving force of the e-government in all of its public institutions will be the application of business management for public administration purposes and the use of modern information technologies for the effective information management and the establishment of improved relations with consumers, partners and suppliers. In addition, (Mitullah & Waema, 2005) advocate for the implementation of E-government strategy within a well-defined and integrated national policy framework which will oversee its objectives *(see Table 2)* being realized. This would be through the public sector reforms under the initiative of Result Based Management (RBM) that is expected to shape organizations and work activities for the achievement of predetermined outputs/results, and the implementation of e-governance strategy in line with Information Communication and Technology (ICT) policy, strategy development and implementation.

**Table 2: E-governance Objectives**

- To increase efficiency and effectiveness, enhance transparency and accountability in the delivery of government services through the use of information technologies.
- Improve the internal workings of government to be externally-oriented and more customer-focused.
- Facilitate collaboration and the sharing of information within and between government agencies.
- Reduce significantly transaction costs leading to savings.
- Encourage participation and empowerment of citizens including the disadvantaged groups and, communities in the rural and remote areas through closer interaction with the government.
- Attract foreign investments by providing faster access to information and government services.

Adopted from (Mitullah & Waema, 2005)

## Information Communication Technologies and Efficient Service Delivery in Local Government

Effective technologies have contributed greatly in the shift from low income level country to middle income level especially in the management of the factors of production (Porter, Sachs, Cornelius, McArthur, & Schwab, 2002). In a devolved government, effective adoption and application of technologies at the counties level will have positive synergetic effect to the overall governance structure.

The current endeavors in the Kenya Local authorities e-Governance framework focuses on implementation of the following three categories of management systems in the individual local governments (Ministry of Local Government Kenya).

1) Local Authorities Integrated Financial Operations Management System (LAIFOMS) which provides three instrumental facilities i.e. Revenue collection facility, expenses management facility and budget preparation – monitoring facility.
2) Local Authorities geographical information system (LAGIS) that focuses on managing map and geo-reference properties with a view to determine the potential for collection of rates from property owners within a given in local authority.
3) Local Authorities human resource management systems (LAHRMS). That manages all aspects of human capital within a local government.

## Background study of Workflow

### What is workflow and workflow management systems?

There are numerous working definitions of workflow a number of authors in the field of workflow systems have defined workflow in different ways. For example, according to Workflow Management Coalition (Coalition, 1994) defines workflow as the computerized facilitation or automation of a business process, in whole or part.

Hollingsworth describes workflow as the execution and automation of business processes where tasks, information or documents are passed from one participant to another for action, according to a set of procedural rules (Hollingsworth, 1995). Center for Technology in Government (CTG) describes workflow as the movement of documents and tasks through a business process which can be a sequential progression of work activities or a complex set of processes each taking place concurrently, eventually impacting each other according to a set of rules, routes, and roles (Government, 1997). Hollowitz and Wilson defines workflow as the automation of a business process, in whole or part, during which documents, information or tasks are passed from one participant to another for action, according to a set of procedural rules (Wilson, March 2003).

In general the above definitions highlight a consensus among the authors on three issues namely; automation of business processes, movement of documents, information or tasks from one participant to another for action, and a set of procedural rules. The working definition used by the authors of this paper is the one coined by (Wilson, March 2003):

Workflow management systems (WFMs) are a system that completely define, manages and executes "workflows" through the execution of software whose order of execution is driven by a computer representation of the workflow logic (Coalition, 1994). They are tools that are used to define, manage and execute workflows on computing resources. They are tools used by organizations to increase their competitive advantage, customer service, productivity, and conformity with standards (Jorge Cardoso, Robert P. Bostrom, & Amit Sheth, 2006) .

**Workflow Evolution and model**

Workflow Management Coalition (WFMC) is a global institution of developers, adopters, analysts, consultants and university research groups whose aim is to develop standards that will be of help to investors of workflow systems especially when these systems are required to interoperate with those of other organizations whenever business processes are conducted across organizational boundaries. The evolution of workflow has taken shape from a number of product areas which includes: image processing, document management, electronic mail and directories, groupware applications, transaction based transactions, project support software and business process reengineering and structured system design Tools (Hollingsworth, 1995).

The workflow model in Figure 1 below has been developed using the generic workflow application architecture (see *Appendix A*) through interfaces that enable products to interoperate at different levels. The workflow model presents five major interfaces whose functions can be summarized in Table 1.

Table 1: Workflow Functional areas

| Functional Area | Purpose |
|---|---|
| Workflow Enactment Service | - Create, manage , and execute workflow instances<br>- Enables one workflow system to pass work item seamlessly to another workflow system |
| Process definition | - Used to analyse, model, describe and document business processes<br>- Enables to use process definition as an input to the workflow engine |
| Workflow client application | - Provides an interface between the client and workflow engine through worklists- (queue of work items assigned to a particular user). The worklist may have different active instances of a single process or individual items resulting from activation of several different processes |
| Invoked application functions | - Defines the interface for calling an application for handling an activity.<br>- Provides communication between the workflow systems and the all potential applications which might exist in an heterogeneous product environment |
| Administration tool | - Helps in managing the operations of the users, roles, resource control, process supervisory functions, and process status functions |

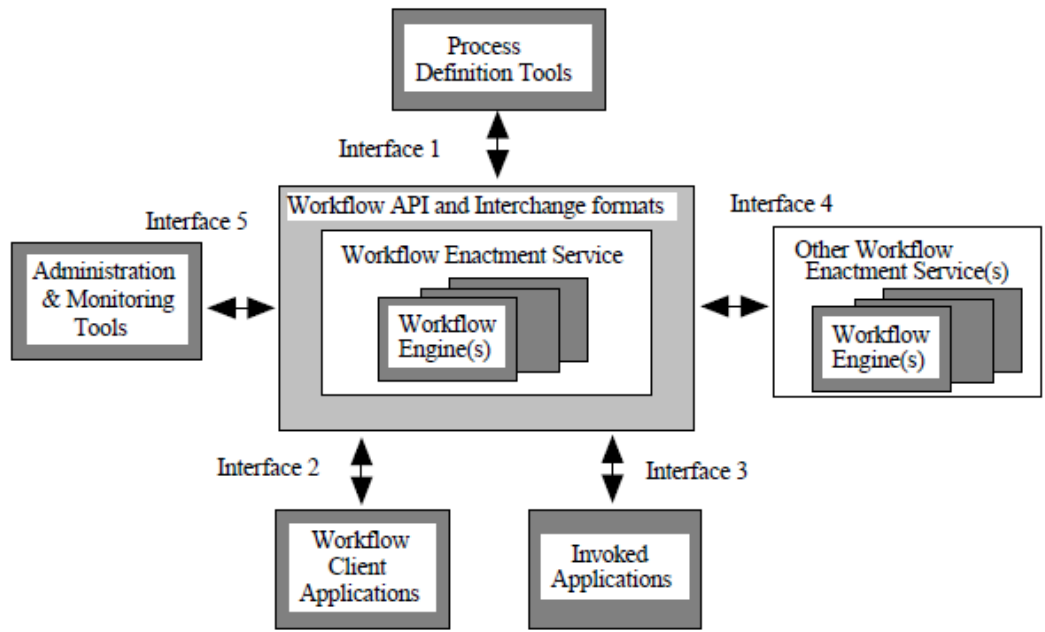

Figure 1: Workflow Model components and interfaces

Adopted from Hollingsworth (1994)

**Application of workflow management systems in Local authorities**

Workflow management system can be adopted in the existing e- Governance framework to provide a link that synchronizes the inputs and out puts of the three and other systems to be introduced e.g. workflow systems can monitor a citizens request e.g. (renewal of a license) by mapping it to the officer providing the service (the details about the officer are provided by LAHRMS ), the citizen is able track the task execution and get notification , after the service the billing is taken by LAIFOMS. The managers can also be able to track performance of their officers. This can help to track service provision and realization of acceptable quality service level. The current systems do not provide this link that extends to the citizens domain and other Government agencies through secure web based connection. Developing extension provided by the workflow system requires introduction of good knowledge management practices to ensure transfer of domain experiences (Tacit Knowledge, Explicit and Implicit Knowledge) which are currently informal and not well documented in public service in Kenya. A generic proposed framework for Knowledge Management in Software Engineering in East Africa (Karanja Evanson Mwangi, Lawrence Xavier Thuku, & karanja, 2010) can be adapted (see Appendix B) .

The Figure 2 below shows the adoption and integration of work flow systems in the existing Framework.

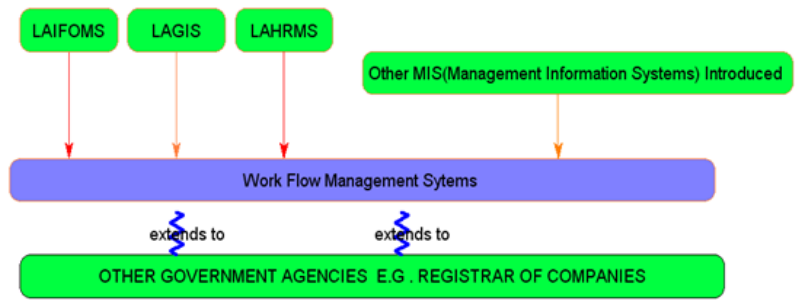

Figure2 : Adoption of workflow in the existing E-government strategy in Local Authorities in Kenya

The success of Electronic government in local government is anchored on effective governance process, resources Management, active citizens' participation and social auditing (Bridge.org; Karanja, 2009; Soriano, 2007; Sturges, 2004).

**Workflow architecture for implementing workflow systems in the Local Authorities of Kenya**

The success for the workflow adoption in the local authorities in Kenya can be based on implementing the workflow architecture as indicated in **Figure 3** below, the background on functionalities of this framework is based on the windows workflow foundation (foundation, 2010). The framework is based on four main functionalities which are integrated with other business process management namely:

1) Task Management Service
2) Tracking Service
3) Notification service
4) Identity service
5) Interception with other business process management

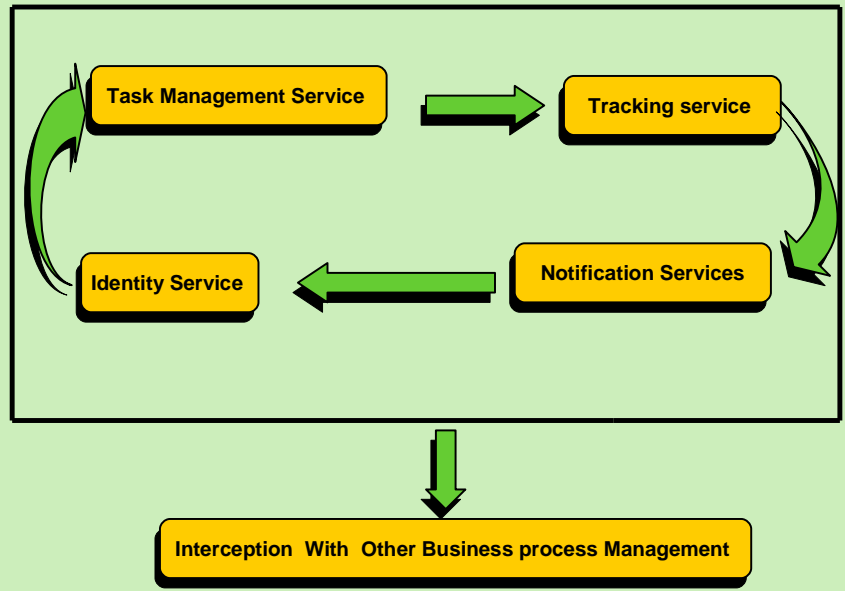

Figure 3: Workflow Architecture: Workflow Usage in Local Autorities based on functionalities

In the next sub-sections, we discuss the granules that make up each of the stages and highlight its building blocks of achieving a comprehensive usage of workflow systems in the local authorities:

1) **Task Management Service:**

Tasks form the communication unit between the business processes and people. For example, the supervisor in a purchasing department needs to approve the request of purchasing items in the local authority. Once the user enters the required data for purchasing the items the tasks constantly switch from one state to another during the workflow lifetime, the states describes the task life cycle and include *pending*, where the task has been created; claimed, where a user has *claimed* the task and has received its input data; *completed*, where a user has finished the task and provided its output data; and *failed*, where a user has finished the task and provided a fault message.

Tasks in the workflow systems are normally associated with time frames: *expiration, escalation, delegation, and renewal*. In our example above, the approval task can *expire* if the supervisor does not act on it in a specified period of time. This expired task can be subsequently *escalated* to another course of action or assignment. Also, the supervisor can decide to *delegate* the task to another person (a manager, for example) to act in his or her place. The manager can also decide to *necessitate* another manager to gather additional insight. If this second manager does not act on the task in a given time frame, the task will be *renewed* for another period of time.

2) **Tracking Service**

The purpose of the tracking service is to ensure that the process of the task created is tracked by the supervisor in a consistent manner and provides scalability and reliability in the business process. It helps a supervisor to be able to track any event in the event of runtime of the process.

3) **Notification Service**

This service is used to create and dispense the information created during the workflow runtime to various stakeholders such as the citizens, managers of various portfolios, etc through either use of intranet for the local stakeholders, mobile devices, and also through a web based platform.

4) **Identity Service**

The identity service provides a secure authentication process for various users through their properties, roles, privileges and membership.

5) **Interception with other business process management**

To realize effective service delivery, the workflow system should have a seamless connectivity with other supportive systems e.g. in this case the system can extend to other government entities that provide supportive entities such as registration of businesses or persons.

**Benefits of workflow systems in Local authorities**

In this section we highlight broad benefits that adoption of workflow systems in local authorities can realize:

1) Resource Management efficiency:- Counties have a prime responsibility of managing vast resources at the grassroots level and extending to help realize the common good for the nation such resources include new asset development, human resources, creation on new information partnerships that leads to new business ventures, consolidation of knowledge and best practices in various areas.

2) Process improvement and control: - Improvement in delivery of general processes that are also denoted as Tasks such includes Business Registration and issuing of licenses or permits can easily be controlled and cycle time of each automated task reduced.

3) Improved quality of customer service:- To realize their set service ratios and increase revenue collection, adoption of workflow architecture (figure 2) which extends to giving status reports and notification can be extended to give feedback to customers at their own convenience e.g. through mobile devices.

4) Business process improvement: - the core business of a county is to provide efficient and quality services to the customers (Citizens) in line with the above benefits.

**Conclusion and areas of further research**

The paper begins with an introduction to the devolved government structure in Kenya with an emphasis on formation and working of the counties. A brief overview of specific articles of the constitution of Kenya that explores the role of counties and economic blueprint vision 2030 is expounded. Exploratory literature study on workflow system is conducted; how they work and their functional area. We evaluate the effective integration of workflow systems in the current local authorities E- government strategy in Kenya and propose a generic framework for adoption of workflow system. The adoption of workflow systems aims to map various functions carried out by the counties to realize the benefits highlighted. For effective application of this framework, the authors suggest that further research be undertaken to determine the following:

1) The Criteria for baseline standards (means of determining basic acceptability level) in the knowledge management process as on what is to be assimilated from the framework (Appendix B: Proposed hybrid framework for software development). An empirical study can be conducted.

2) The most effective system development approach to be used, the tradeoff between adoption of an off the shelf system and Bespoke development.

3) The viability of free and open source technologies in development and integration of workflow systems in the counties. The authors recommend the investigation to extend to the assessment of benefit of incorporating local natural languages in the citizens' access points.

**References**

Bridge.org. **http://www.bridges.org/about**

Coalition, W. M. (1994). *The Workflow Management Coalition Specification*.
foundation, W. W. (2010). Windows workflow foundation services.   Retrieved 25th July, 2010, from http://msdn.microsoft.com/en-us/library/ms735887(VS.90).aspx

Government, C. f. T. i. (1997). *Introduction to Workflow Management Systems*  University at Albany / SUNY.

Hollingsworth, D. (1995). *The Workflow Reference Model* (No. TC00-1003): Workflow Management Coalition.

Jorge Cardoso, Robert P. Bostrom, & Amit Sheth. (2006). Workflow Management Systems vs. ERP Systems: Differences, Commonalities, and Applications *International Journal of Computer  science*

KAM, s. (2009). Radical Changes recommended for Local Authorities. Retrieved 20[th] july 2010, from http://www.kam.co.ke/?itemId=17&newsId=177

Karanja  Evanson  Mwangi, Lawrence Xavier Thuku, & karanja, J. P. K. (2010, 25[th]-26[th] August ). *Software Development Industry In East Africa: Knowledge Management Perspective And Value Proposition.* Paper presented at the African International Business And Management (AIBUMA) ConferenceE Kenyatta International Conference Centre (KICC) Nairobi, Kenya

Karanja, E. M. (2009, 4[th] - 5[th] September ). *ICT As An Engine For Sustainable Growth And Development: The Role And Opportunities For The Local Authorities In Kenya.* Paper presented at the 10[th] Annual Strathmore University ICT Conference, Nairobi Kenya.
Kenya, R. o. (1999). *Report of the rationalization and staff rightsizing for effective operation of the Ministry of Local Government*. Nairobi.
The Constitution of Kenya (2010).
Ministry of Local Government  kenya. *Information, education and communication Strategy for the ministry of local government December 2008*. http://www.localgovernment.go.ke/index.php?option=com_jdownloads&Itemid=49&task=view.download&cid=1.

Mitullah , & Waema, T. (2005). *State of ICT and Local Governance, Needs Analysis and Research Priorities*. Paper presented at the Local Governance and ICTs Research Network for Africa

NESC, T. N. E. S. C. *Kenya Vision 2030: Transforming National Development.* . Retrieved 25[th] July 2010. from http://www.nesc.go.ke/News&Events/KenyaVision2030Intro.htm.

Porter, M. E., Sachs, J. D., Cornelius, P. K., McArthur, J. W., & Schwab, K. (2002). *Executive summary: competitiveness and stages of economic development*. New york.

Soriano, C. (2007). Exploring the ICT and Rural Poverty Reduction Link : Community Telecentre and Rural Livelihoods in Wu'an China. *The Electronic Journal on Information Systems in Developing Countries EJISDC, 1*(1), 1-15.

Sturges, P. (2004). Corruption, Transparency and a Role for ICT? . *International Journal of Information Ethics : , 2(11)*.

Wilson, J. H. a. C. (March 2003). The market for IT solutions. *Journal of Technology research institute*.

# Appendix A

## Figure 4: Generic workflow product structure

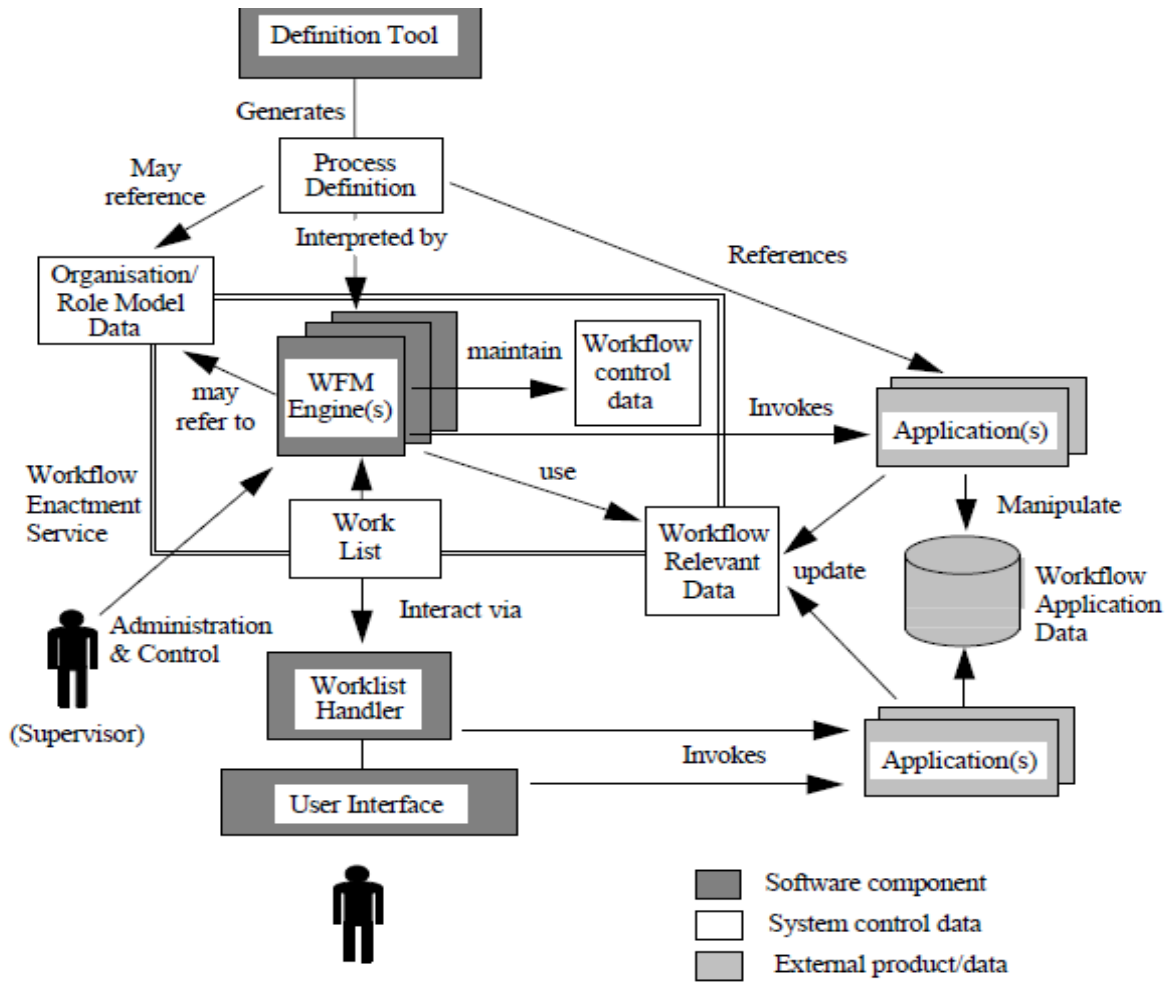

Adopted from (Hollingsworth, 1995)

# Appendix B: Proposed hybrid framework for software development

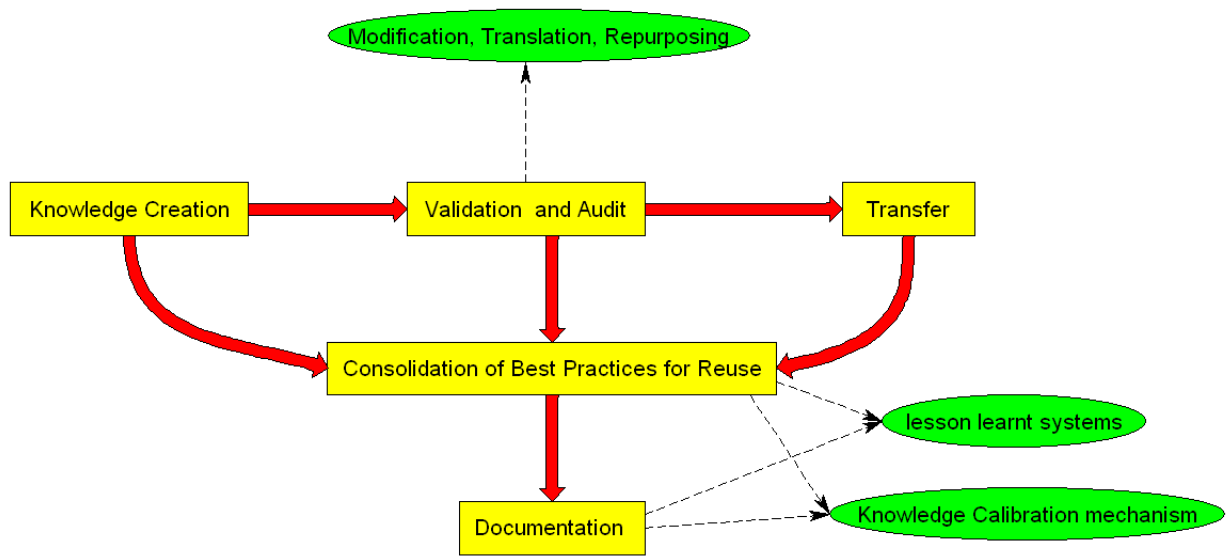

Hybrid Framework for KM in Software Development

**ABOUT THE AUTHORS**

**Lawrence Xavier Waweru Thuku**: Thuku has Master of Science in Information Technology (Msc. IT) from Strathmore University and various ICT industry Certifications. He is a Lecturer and a consultant in the Information Technology field. His research interests include Workflow Management systems, Decision Support systems, Security in systems and knowledge management.

**Karanja Evanson holds an Msc in computer science (Makerere University) with a bias in machine learning and security, among other qualifications. He is a Lecturer and an entrepreneur. His research interests includes, programming languages, E- government, knowledge management, Information security and Workflow Management systems.**